\documentclass[twocolumn,a4paper,pre,superscriptaddress,showpacs,nofootinbib,citeautoscript,floatfix]{revtex4}

\usepackage{bm}
\usepackage{graphicx}
\usepackage{amsmath}
\usepackage{amssymb}

\DeclareMathOperator{\sgn}{sgn}
\newcommand{\ham}{\mathcal{H}}
\newcommand*{\ket}[1]{\left \lvert {#1} \right \rangle}
\newcommand*{\bra}[1]{\left \langle {#1} \right \rvert}
\newcommand{\way}[1]{^{\text{#1}}}

\begin{document}
\title{Quantized charge pumping by surface acoustic waves in ballistic quasi-1D channels}

\author{Vyacheslavs Kashcheyevs}
\email[e-mail: ]{slava@latnet.lv}
\affiliation{School of Physics and Astronomy, Raymond and
Beverly Sackler faculty of Exact Sciences, \\
Tel Aviv University, Tel Aviv 69978, Israel}

\author{Amnon Aharony}
\affiliation{School of Physics and Astronomy, Raymond and
Beverly Sackler faculty of Exact Sciences, \\
Tel Aviv University, Tel Aviv 69978, Israel}

\affiliation{Department of Physics, Ben Gurion University,
Beer Sheva 84105, Israel}

\author{Ora Entin-Wohlman}
\affiliation{School of Physics and Astronomy, Raymond and
Beverly Sackler faculty of Exact Sciences, \\
Tel Aviv University, Tel Aviv 69978, Israel}
\affiliation{Department of Physics, Ben Gurion University,
Beer Sheva 84105, Israel}

\affiliation{Albert Einstein
Minerva Center for Theoretical Physics \\ at the Weizmann
Institute of Science,  Rehovot 76100, Israel}

\date{February 24, 2004}

\begin{abstract}
Adiabatic pumping of electrons induced by surface acoustic waves
(SAWs) in a ballistic quasi-1D quantum channel is considered using
an exactly solvable tight-binding model for non-interacting
electrons. The single-electron degrees of freedom, responsible for
acoustoelectric current quantization, are related to the
transmission resonances. We study  the influence of experimentally
controllable parameters (SAW power, gate voltage, source-drain
bias, amplitude and phase of a secondary SAW beam) on the
plateau-like structure of the acoustoelectric current. The results
are consistent with existing experimental observations.
\end{abstract}


\maketitle

\section{Introduction}
Single electron transport through low-dimensional mesoscopic
structures, driven by surface acoustic waves (SAWs), is a subject
of active experimental \cite{Shilton96,Talyanskii97,Talyanskii98,
Cunningham99,Cunningham00,Ebbecke00,Janssen01,Robinson02,
Fletcher03} and theoretical
\cite{Aizin98,Gumbs99,Flensberg99,Maksym00,
Galperin01,Robinson01,Aharony02PRL,Margulis02,Entin03japan}
research, with potential applications in metrology
\cite{Flowers01} and new computation technologies \cite{Barnes00}.
In a typical experimental setup, a quasi-one dimensional ballistic
channel is defined in a AlGaAs/GaAs heterostructure and a SAW is
launched in the longitudinal direction at a frequency $\omega/2
\pi$ of several GHz. Under appropriate conditions, the
acoustoelectric dc current $I$ exhibits a staircase plateau-like
structure as function of the gate voltage (which controls the
depletion of the channel) and of the SAW power. At the plateaus,
the current saturates at quantized values $I= e (\omega/ 2\pi) m$,
corresponding to the transfer of an integer number $m$ of
electrons per each period of the SAW (here $e$ is the electron
charge). The first plateau is the most flat and robust to changes
in the control parameters; the higher plateaus become less and
less pronounced as the plateau number $m$ is increased. In
addition, the effect of factors such as source-drain bias
\cite{Shilton96,Talyanskii97,Cunningham00}, temperature
\cite{Shilton96,Janssen01}, gate geometry \cite{Cunningham99}, a
secondary SAW beam \cite{Talyanskii97,Cunningham99}, and
perpendicular magnetic field \cite{Cunningham00} on the staircase
structure and the quality of the first plateau have been studied
experimentally.

In the experiment, the plateaus are observed below the conductance
pinch-off, when electrons in the source and in the drain
reservoirs are separated by a potential barrier. This observation
forms the basis for the simple qualitative explanation of the
quantized transport which has been proposed in the first
experimental report \cite{Shilton96} and further refined in Refs.\
~\onlinecite{Aizin98,Cunningham99}, and ~\onlinecite{Gumbs99}.
They argue that when the wavelength $\lambda$ of the SAW is
comparable with the size of the depleted region $L$ (as it is in
the experiments  \cite{Shilton96,Talyanskii97,Cunningham99}), a
single potential well forms on top of the static barrier. This
potential well then acts as a dynamic quantum dot, which can hold
an integer number of electrons due to the Coulomb blockade effect.
The captured electrons are transferred from one side of the
barrier to the other, with possible quantization errors due to
back-tunnelling \cite{Aizin98,Gumbs99}. In the above description,
the formation of the quantum dot and the transport of the
localized electrons are treated separately. Particular effects
which have been studied theoretically within this picture are the
non-adiabatic effects at the quantum dot's formation stage
\cite{Flensberg99}, and the classical dynamics of the already
confined interacting electrons \cite{Robinson01}.

A different perspective on the problem has been suggested in
Refs.\ ~\onlinecite{Aharony02PRL} and ~\onlinecite{Levinson00PRL}.
This approach relates the acoustoelectric transport to adiabatic
quantum pumping of non-interacting electrons. The external
potential, generated by the SAWs and by the control gates, is
viewed as a perturbation acting on a coherent quantum wire
\cite{Entin02form}. The resulting  ``staircase'' structure of the
acoustoelectric current and its dependence on model parameters
within this approach  have been studied in Refs.\
~\onlinecite{Aharony02PRL} and ~\onlinecite{Entin03japan}, using
the adiabatic approximation in conjunction with an exactly
solvable one-dimensional (1D) tight-binding model. This theory
yields a crossover from a non-quantized acoustoelectric transport
to the quantized limit as the SAW power and/or the static barrier
height are increased. Although this picture requires  the Coulomb
interaction in order to set  the energy scale of the problem
\cite{Gumbs99,Flensberg99,Robinson01}, the main qualitative
features of the experiment can be reproduced within a model of
non-interacting spinless electrons \cite{Maksym00,Aharony02PRL}.

In this paper we extend the results of Ref.\
~\onlinecite{Aharony02PRL}. The mechanism of quantized transport
is elucidated by using a resonance approximation for adiabatic
pumping \cite{VKAAOE03res}. Both current quantization and
transmission resonances are determined by the quasi-bound states
of the electrons captured by a moving potential well. New effects,
including the influence of a counter propagating SAW, static
potential asymmetry and source-drain bias on the number and shape
of the quantization steps, are considered. We compare our
qualitative conclusions with the published theoretical and
experimental results. In particular, tuning the amplitude and the
phase of a weak secondary SAW is found to improve the quantization
by accordance with an earlier experimental report
\cite{Cunningham99}. For this effect, we propose a new
quantitative relation between the phase and the amplitude of the
optimal secondary SAW which can be easily checked using existing
experimental setups.

The results are presented as follows. In Section~\ref{SecModel},
we describe the model \cite{Aharony02PRL} and the algorithms for
calculating the adiabatic current.  In Section~\ref{SecSteps}, we
explain the formation of the integer plateaus and make
quantitative analytic estimates by applying the resonance
approximation \cite{VKAAOE03res} to the model of Ref.\
~\onlinecite{Aharony02PRL}. Building on these results we analyze
in Sec.~\ref{SecPerturbations} additional factors, not described
previously, such as reflected SAWs, source-drain bias and gradual
screening of the pumping potential. Finally, a discussion of our
results in the context of related work is presented in
Sec.~\ref{SecDiscussion}, together with several conclusions.

\section{The Model} \label{SecModel}
\subsection{The tight-binding Hamiltonian}
\begin{figure}[tb]
  \includegraphics[clip=true,width=8.6cm]{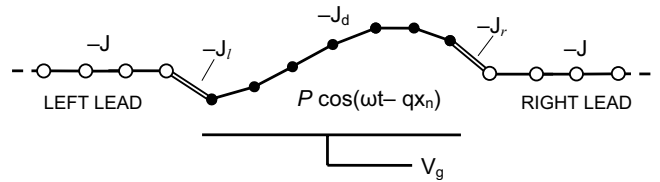}
  \caption{One-dimensional tight binding model for
  SAW-induced pumping.}\label{fig0}
\end{figure}

The model consists of a nanostructure and two 1D leads connecting
its ends to the electronic reservoirs (Fig.~\ref{fig0}). The leads
are modelled by two chains of sites with vanishing on-site
energies and nearest-neighbors hopping amplitudes $-J$. An
electron moving in the lead has the energy $E(k)=-2 J \cos ka$,
where $k$ is the wave vector and $a$ is the lattice constant.

The nanostructure  represents the SAW-affected part of the
quasi-1D channel. It is described by $N$ tight-binding sites $n=1,
2, \ldots N$, that form a chain of length $L=(N-1)a$. The non-zero
elements of its Hamiltonian matrix, $\ham_0(t)$, are constant
nearest-neighbors hopping amplitudes $-J_d$ and diagonal
time-dependent on-site energies $\epsilon_n(t)$. The connection
between the ideal leads and the time-dependent part of the
constriction is introduced through a hopping amplitude $-J_l$
($-J_r$) between the left (right) lead and the site $1$ ($N$) of
the nanostructure. The resulting full Hamiltonian of the quantum
wire will be denoted $\ham$. In the trivial case where
$J_l=J_r=J_d=J$ and $\epsilon_n=0$, $\ham$ describes an ideal 1D
tight-binding lattice.
The time-dependence of the on-site energies $\epsilon_n(t)$
reflects the SAW-induced piezoelectric potential, which is
responsible for the pumping effect. The simplest case is that of a
single running wave as introduced in Ref.\
~\onlinecite{Aharony02PRL},
\begin{gather} \label{potential_original}
   \epsilon_{n}(t)  = -V_g+
    P \, \cos \left( \omega  t - q x_n  \right) \, .
\end{gather}
Here, $V_g$ is the gate voltage (in energy units), $q$ is the
wave-vector of the SAW, and $P$ is the amplitude of the
piezoelectric potential, induced by a SAW running from left to
right (for $q>0$). The origin of the spatial coordinate
$x_n=(n-1)a - L/2$ is chosen to be at the middle of the channel.

\subsection{The acoustoelectric current} \label{FullCalc}
The adiabatically-pumped current flowing between two reservoirs
with equal electrochemical potentials $\mu_l=\mu_r$, is usually
calculated using the Brouwer formula \cite{BPT94,Brouwer98}. We
use an equivalent formalism, developed in Ref.\
~\onlinecite{Entin02form}, which also includes the effects of a
finite  bias $e V_{SD} \equiv \mu_l -\mu_r \not =0$. The total
instantaneous current, $I_{\alpha}(t)$, of spinless electrons from
the lead $\alpha= l, r$ into the nanostructure consists of two
parts, $I_{\alpha}(t)=I_{\alpha}\way{pump}+ I_{\alpha}\way{bias}$.
These two parts can be conveniently written down using the
instantaneous scattering states $\ket{\Psi_\alpha(t)}$ (normalized
to a unit flux), the instantaneous transmission coefficient
$\mathcal{T}(t)$ and the overall scattering phase $\theta(t)$ of
the nanostructure, \cite{Entin02form}
\begin{gather}
I_{\alpha}\way{pump}  = \frac{e}{4 \pi \hbar} \int   d  E\,
\bra{\Psi_\alpha(t)} \dot{\ham} \ket{\Psi_\alpha(t)}
\frac{\partial (f_l+f_r)}{\partial E} \, , \label{pump_current} \\
\begin{split}
I_{\alpha}\way{bias}  = \frac{e}{2 \pi \hbar} \int d  E\,
\Bigl \{  & (f_l-f_r) \,  \mathcal{T}   +
\frac{\hbar}{2} \frac{\partial (f_l-f_r)}{\partial E} \, \mathcal{T} \, \dot{\theta}
\Bigr \}  \, . \label{bias_current}
\end{split}
\end{gather}
Here $f_{\alpha}(E)=1/[1+e^{\beta (E-\mu_{\alpha})}]$ is the Fermi
distribution with  $\beta=1/k_B T$ ($T$ is the temperature). If
the system is unbiased, then $I_{\alpha}\way{bias}=0$ and
Eq.~\eqref{pump_current} can be shown to
reproduce \cite{Entin02form,VKAAOE03res} the Brouwer
formula \cite{BPT94,Brouwer98}. On the other extreme, if no pumping
potential is applied, $I_{\alpha}\way{pump}=0$ and
Eq.~\eqref{bias_current} leads to the Landauer
formula \cite{Landauer70} for the conductance,  $G=(e^2/h)
\mathcal{T}$.

For most of the discussion we assume both the bias voltage
$V_{SD}$ and the temperature $T$ to be zero. In this case only
electrons at the Fermi energy $\mu_l=\mu_r=E_F$ participate in the
scattering. Solving the scattering problem for the potential
\eqref{potential_original} and using Eq.~\eqref{pump_current}
yields the charge $Q$ pumped over one period (the average dc
component of the current)  \cite{Aharony02PRL},
\begin{align}\label{chargeexact}
   Q = \int_0^{2 \pi/\omega} \!\!\!\!
   dt \, I^\text{pump}_l(t) =
  \frac{e \tilde{J_l} \sin ka}{\pi}
   \int_{0}^{2 \pi/\omega}
   \!\!\!\! dt \sum_{n=1}^N \dot{\epsilon}_{n}|g_{n,1}|^{2} \, , \\
  \left [g^{-1}\right ]_{n,n'} =
  \left [E \bm{I}- \ham _0 \right ]_{n,n'} 
  +  \delta_{n,n'} \, e^{ika} \left(
  \delta_{n,1} \tilde{J}_{l}+\delta_{n,N} \tilde{J}_{r} \right ) \, ,
  \label{green}
\end{align}
where $\tilde{J}_{l;r} \equiv J^2_{l;r}/J$ and $k$ is the Fermi
wavenumber, $E_F\equiv E(k)$. The instantaneous transmission is
\begin{equation} \label{transmission}
  \mathcal{T}(t) = 4 |g_{N,1}|^2 \, \tilde{J}_l  \tilde{J}_r \, \sin^2 ka \, .
\end{equation}

The integrand in Eq.~\eqref{chargeexact} is a meromorphic function
of $z =\exp( i \omega t)$, with $2 N$ pairs of complex conjugate
poles. Therefore, the integration of Eq.~\eqref{chargeexact} may
be carried out exactly, once the positions of the poles are
determined by solving numerically the corresponding algebraic
equation of degree $2 N$.

\subsection{Resonance approximation} \label{ResonanceCalc}
The second term on the r.h.s. of Eq.~\eqref{green} is  the
self-energy addition to the Green's function of the isolated
channel, due to the coupling to the external leads. When the
latter is sufficiently small,  the total pumped charge can be
divided into contributions from separate single-particle levels of
$\ham_0$. A systematic development of this approach leads to the
resonance approximation for pumping, which is discussed in detail
in Ref.\ ~\onlinecite{VKAAOE03res}. Here we summarize the
resulting algorithm for calculating the pumped charge in the this
approximation.
\begin{enumerate}
\item  Solve the instantaneous eigenvalue problem $\sum_{n'}\left
[\ham_0\right ]_{n,n'} \psi_{n'}^{(m)} = E_m \, \psi_{n}^{(m)}$
and obtain the approximate resonance energies $E_m(t)$.

\item Calculate the time-dependent decay widths of each resonance
into each lead,
\begin{equation} \label{decayWidth}
 \left \{ \Gamma^{(m)}_{l}, \Gamma^{(m)}_r \right \} = \left \{
 \tilde{J_{l}} \bigl \lvert \psi_{1}^{(m)}  \bigr \rvert^2,
 \tilde{J_{r}} \bigr \lvert \psi_{N}^{(m)} \bigr \rvert^2 \right\}
 \sin ka \, .
\end{equation}

\item For each $m$, find all such times $t_{m,j}$ at which the
resonance condition $E_m(t_{m,j})=E_F$ is satisfied.

\item At each resonance time $t=t_{m,j}$, compute the partial
charge transferred between the left lead and the $m$th quasibound
state in the channel,
\begin{equation}\label{Qrespartial}
 \Delta Q_{m,j} =  \left . \frac{e \,
 \Gamma_l^{(m)}}{\Gamma_l^{(m)} +\Gamma_r^{(m)}} \, \right
 |_{t=t_{m,j}} \, .
 \end{equation}

\item Calculate the total charge pumped from left to
right\footnote{Due to charge conservation, it is sufficient to
calculate the charge transfer from the left reservoir. Therefore,
the channel index $\alpha$ is fixed to $\alpha=l$ in
Eqs.~\eqref{Qrespartial} and  \eqref{Qtot}.}:
\begin{equation} \label{Qtot}
Q^{\text{res}} =-\sum_{m,j} \Delta Q_{m,j} \sgn \dot{E}_m(t_{m,j}) \, ,
\end{equation}
or set $Q^{\text{res}}=0$ if no resonances were found in step~3.

\end{enumerate}

The algorithm has a direct physical interpretation
\cite{VKAAOE03res}. Whenever the energy $E_m$ of a (quasi-)bound
state crosses the Fermi level $E_F$, an electron either occupies
(``loading'') or leaves (``unloading'') this state. The
corresponding unit pulse of current is distributed between the
channels proportionally to the $\Gamma_{\alpha}^{(m)}$'s. Except
for specifically designed Hamiltonians $\ham_0(t)$,
$Q^{\text{res}} \to Q$ in the limit of vanishing couplings
$\Gamma_{\alpha}^{(m)} \to 0$.

The resonance approximation fails when either (i) the total width
of a particular resonance is larger than the distance to the next
energy level; or (ii) the partial decay widths
$\Gamma_{l;r}^{(m)}$ change considerably while the system is at
resonance \cite{VKAAOE03res}. As discussed in detail in the
following section, these restrictions  become significant for the
non-quantized transport, but have little influence on the shape of
the current quantization steps. In all the cases in which the
resonance approximation is inadequate, we rely on the results of
an exact calculation.

\section{Formation of quantization steps} \label{SecSteps}
\subsection{Application of the resonance approximation}
The results of a full calculation (as outlined in
Sec.~\ref{FullCalc}) show \cite{Aharony02PRL} that the pumped
charge, $Q$, follows a staircase-type dependence on the gate
voltage, $V_g$, and/or on the SAW amplitude, $P$, for a wide range
of the model parameters. This  `quantization' can be related to
the structure of the transmission resonances
 \cite{Entin02res,Levinson02PhA,VKAAOE03res}. We first establish
this relation quantitatively and then use it to analyze various
aspects of the model.

\begin{figure*}[tbp]
  \includegraphics[width=12cm]{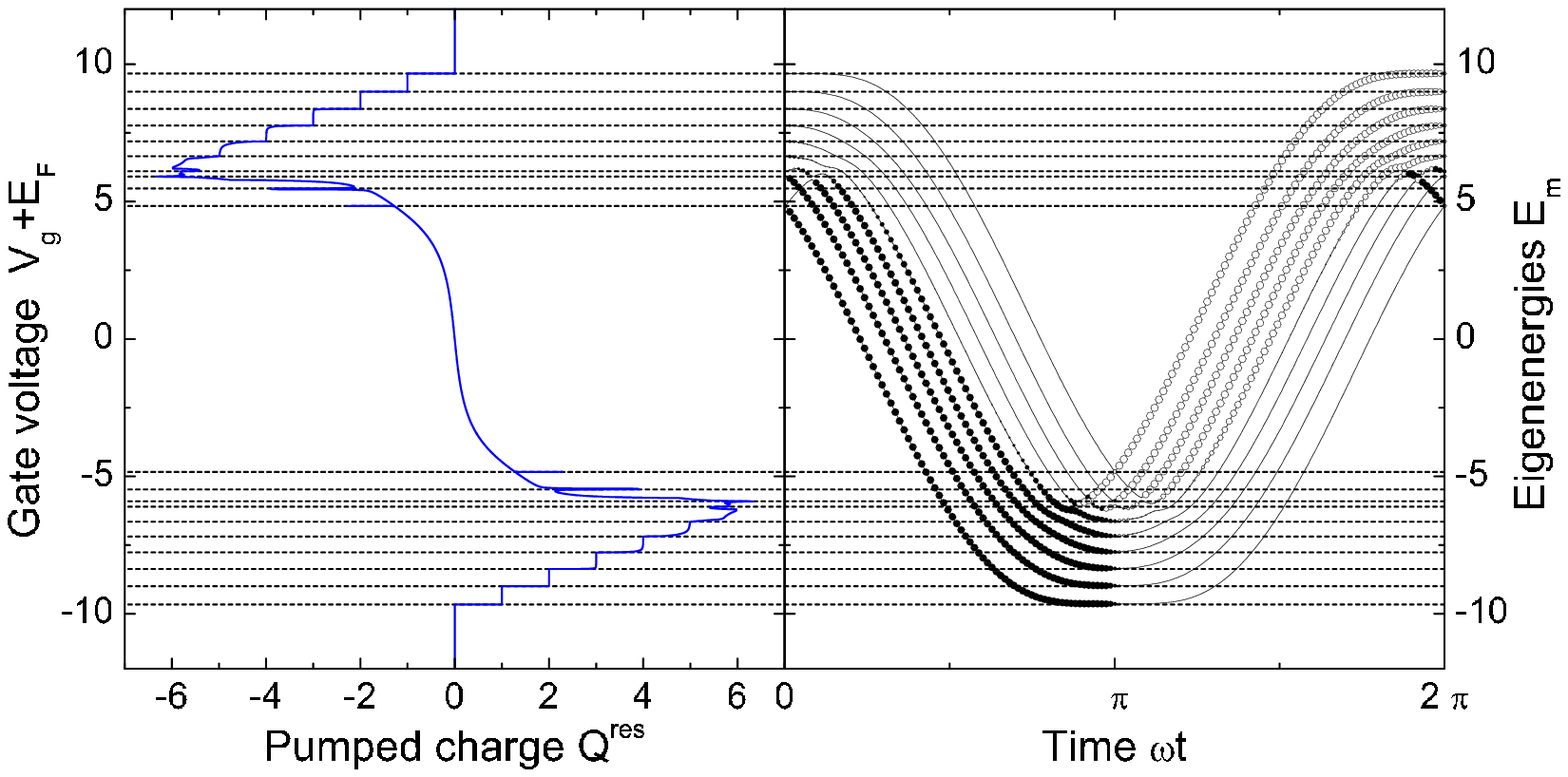}
  \caption{(Color online) Construction of the pumping curve
  $Q^{\text{res}}(V_g)$ in the resonance approximation. Right
  panel: instantaneous energy levels of $\ham_0$
  at $V_g=0$ as  function of time. Left panel: pumped charge
  $Q^{\text{res}}$ (in units of $e$) as  function of gate
  voltage. Horizontal dashed lines show the correspondence between
  sharp features in the pumping curve (left), and the change in the number of
  energy levels at resonance (right); see text for a detailed discussion.
  All energies are given in units of $J_d$; the
  parameters of the potential are: $P=8 J_d$, $\lambda= 4 L$,  $N=10$.\protect\label{fig2}}
\end{figure*}

The calculation of the pumping curve can be visualized using a
diagram like the one shown in Fig.~\ref{fig2}. First, one plots
the instantaneous eigenvalues $E_m$ for $V_g=0$ as  function of
time $\omega t$ (curves in the right panel of Fig.~\ref{fig2}).
The small circles on the top of each curve show the time evolution
of the corresponding partial charge $\Delta Q$: the diameter of
each circle is proportional to $|\Delta Q/e| < 1$; shading is
determined by the sign --- black ($\bullet$) for $\dot{E}_m<0$
(``loading'') and white ($\circ$) for  $\dot{E}_m >0$
(``unloading'') . Once the eigenvalue diagram is constructed, the
set of resonances for each particular value $V_g$ of the gate
voltage is determined graphically: a horizontal line with ordinate
$E_F+V_g$ crosses the eigenvalue curves in the right panel at the
points where the resonance equation $V_g+E_m(V_g=0)=E_F$ is
satisfied (step~3 of the algorithm). The abscissas of the crossing
points determine the resonance times $t_{m,j}$ to be used in
Eqs.~(\ref{Qrespartial}) and ~(\ref{Qtot}). (The dashed horizontal
lines in Fig.~\ref{fig2} mark the extrema of the eigenvalue
curves, and thus correspond to particular values of $V_g$ at which
the number of resonances changes.) Finally, the total pumped
charge, $Q^{\text{res}}(V_g)$, is calculated by summing up the
contributions to Eq.~\eqref{Qtot}: the magnitude and the sign of
each term is given by the small circle at the respective crossing
point in the right panel. The resulting pumping curve
$Q^{\text{res}}(V_g)$ is plotted in the left panel of
Fig.~\ref{fig2}.

Several aspects of the model are illustrated by the construction
in Fig.~\ref{fig2}. One can see that the quantization of the
pumped charge is caused by electronic (hole) states with the
lowest (highest) energy. When resonances occur, (namely, at $\{
t_{m,j}\}$), these states are \emph{localized} near one of the
channel exits --- either $\Gamma_l/ \Gamma_r \ll 1$ or $\Gamma_l
/\Gamma_r \gg 1$ --- and therefore transfer almost integer charges
[Eq.~\eqref{Qrespartial}]. The number of steps counts the number
of localized states involved.

\begin{figure*}[tbp]
  \includegraphics[width=12cm]{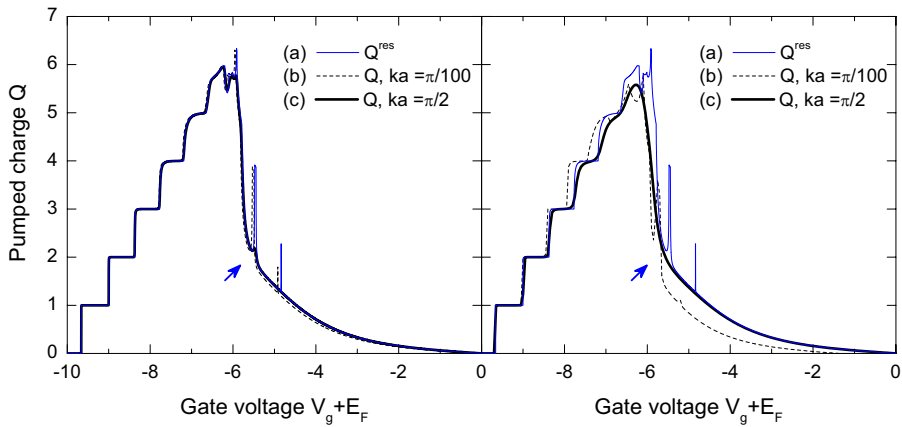}
  \caption{(Color online) Pumped charge versus gate voltage for different
  external parameters: $\tilde{J}=0.16 J_d$ (left panel),
  $\tilde{J}=1 J_d$ (right panel), calculated in the resonance
  approximation (a) and exactly for the bottom of the band (b) and
  at the band center (c). Parameters of the potential are the same
  as in Fig.~\ref{fig2}.\protect\label{fig1}}
\end{figure*}

The exact integration [Eq.~\eqref{chargeexact}] takes into account
the ``external'' parameters of the model, $k a$, $\tilde{J}_l$ and
$\tilde{J}_r$, which are ignored in the resonance approximation.
In the following, we will consider only symmetric couplings,
$\tilde{J}_{l} = \tilde{J}_{r} \equiv \tilde{J}$. We have
calculated the exact pumped charge, $Q(V_g)$, for several values
of the ``external'' parameters, but with the same pumping
potential as in Fig.~\ref{fig2}. Representative results are shown
in Fig.~\ref{fig1} along with the approximate
$Q^{\text{res}}(V_g)$ from Fig.~\ref{fig2} [thin (blue) line]. For
sufficiently small $\tilde{J}$, the exactly calculated curves
contain integer steps and sharp, non-quantized features at large
values of $V_g$ (e.g., the spikes marked by small arrows in
Fig.~\ref{fig1}). The first steps are robust and do not change
their positions as $\tilde{J}$ and $k a$ are varied (except for a
trivial shift of $E_F$). The top of the pumping curve and  the
spikes are more vulnerable: as $\tilde{J}$ is increased, the upper
steps and the sharp features shift and become rounded. Narrow
spikes disappear for $\tilde{J}=J_d$ and $ka$ close to the center
of the band [see curve (c) in the right panel of Fig.~\ref{fig1}].

The resonance approximation reproduces all the details of the
exact calculation for $\tilde{J} \ll J_d$, because the resonance
widths in Eq.~\eqref{decayWidth} vanish in the limit of $\tilde{J}
\to 0$. The non-generic sharp features are determined by the
surroundings  of level anti-crossings (see Fig.~\ref{fig2}), where
the corresponding level spacings are tiny. As we expect form the
validity condition (i) in Sec.~\ref{ResonanceCalc}, the finite
resonance width effects are most important in this region. Indeed,
the discrepancies between the exact and the approximate curves in
Fig.~\ref{fig1} are well correlated with the fact that the shifts
and the widths of the resonance levels for a tight-binding model
are proportional to $\tilde{J} \cos ka$ and $\tilde{J} \sin ka$,
respectively [Eq.~\eqref{decayWidth}].

We have made a similar comparison between the exact integration
and the resonance approximation for several sets of ``internal''
parameters,  $P/J_d$, $\lambda/L$, and $N$. The most important
conclusion is that the stair-case structure of the pumping curve
can be reliably estimated using the resonance approximation.
Hence, we will utilize this useful technique in the following  as
a source for various analytic estimates that will be further
checked versus exact calculations.

\subsection{SAW parameters and the number of quantization steps} \label{subSecNumSteps}
For the lowest part of the spectrum (which is relevant for the
quantized transport), the on-site energies
\eqref{potential_original} can be treated as a potential function
of a continuous spatial coordinate $x_n$. For $\lambda > L$, only
one minimum of this potential can be located inside the
SAW-affected part of the channel. The position of this minimum
$x_0(t)=(t -t_0) v $ moves with the sound velocity $v=\omega/q$
and passes through the middle of the channel at time
$t_0=\omega^{-1} (\pi + 2 \pi \times \text{integer})$. Electronic
states localized in this moving potential well can be approximated
by simple harmonic oscillator wave-functions \cite{Aharony02PRL}.
The corresponding energy spectrum is $E_m= -P-2 J_d + \Delta (m
-1/2)$, $m=1,2, \ldots$, with a constant spacing $\Delta/ J_d =
qa \sqrt{2 P/ J_d}$. The lowest energy wave-function is
approximately a Gaussian,
\begin{align} \label{ground0}
    \psi_n^{(1)}(t) = {\left(\xi^2 \, \pi/2  \right)}^{-1/4}\,
    \exp \{ -[ x_n-x_0(t)]^2/\xi^2 \} \, ,
\end{align}
with  $\xi \equiv 2 a \sqrt{J_d/\Delta}$. The localization length
of the higher levels can be estimated as $\xi_m = \xi \,
\sqrt{m}$.

The harmonic approximation is valid as long as  the wave-packet is
driven adiabatically by a parabolic well and is not perturbed
neither by the ends of the channel, nor by the ``hills'' of the
cosine-shaped potential profile. This implies the validity
condition
\begin{align}
  \xi_m \ll\min[L/2-|x_0(t)|, \lambda/2] \, . \label{condharm}
\end{align}
In order to illustrate the above reasoning, we draw the attention
of the reader to a set of constant and equidistant energy levels
$E_m(t)$ in the right panel of Fig.~\ref{fig2}, in the vicinity of
$\omega t = \pi$. The lowest energy level follows the harmonic
approximation as long as the parabolic minimum is located inside
the channel, that is for the fraction $\lambda/L=1/4$ of the full
period. Higher energy levels remain constant for shorter times,
since their respective localization lengths entering
Eq.~\eqref{condharm} are longer. The harmonic structure of the
energy levels translates into a sequence of equidistant steps in
the pumping curve, $Q(V_g)$, with the same energy spacing
$\Delta$, as shown in the left panel of Fig.~\ref{fig2}. At each
value of gate voltage, $V_g^{(m)} = E_m - E_F$, a new pair of
resonances and another step in the pumping curve emerge. The
plateaus are rather flat because the resonant states at the
loading (unloading) moments are well localized at the entrance
(exit) of the channel.

The number of quantization steps, $N_{\text{steps}}$, is limited
by two competing mechanisms. The first limit is set by the number,
$N_{1}$, of localized states that can be transferred below the
Fermi energy. If $x_n$ can be considered as continuous, the
localization condition is roughly the same as the validity
condition \eqref{condharm} for the harmonic approximation. For $L
< \lambda$ it follows from $\xi_{N_{1}} = L/2$ that $N_{1}=L^2
\Delta/(16J_d \, a^2 )=(\pi \sqrt{2}/8) \, N (L/\lambda)
\sqrt{P/J_d}$. On the other hand, for large enough $P$ the
discreteness of the tight-binding grid cannot be neglected. For a
rough estimate, we assume that the continuous approximation breaks
down if it yields an average distance $\xi_m/m$ between the
successive zeros of the $m$th wave-function, which is smaller than
the inter-site spacing $a$. This happens for $m > N_2$, where
$N_2=N^2/(4 N_1)$. Putting the two limits together we estimate the
number of quantization steps, $N_{\text{steps}}$, as the integer
closest to $\min(N_1,N_2)$. By adjusting the parameters one can
obtain at best a sequence of $N/2$ steps. The optimal parameters
$L=0.3\lambda$, $N=6$, $P=8 J_d$ of Ref.\
~\onlinecite{Aharony02PRL} indeed yield $N_{\text{steps}} \approx
N_1 = 2.83 \approx N_2 \approx N/2$. The decrease in the number of
steps with increasing $L/\lambda$ reported in Ref.\
~\onlinecite{Aharony02PRL} corresponds to the tight-binding
limited regime $N_{\text{steps}} \approx N_2 \propto \lambda/L$.

\begin{figure}
  \includegraphics[width=8cm]{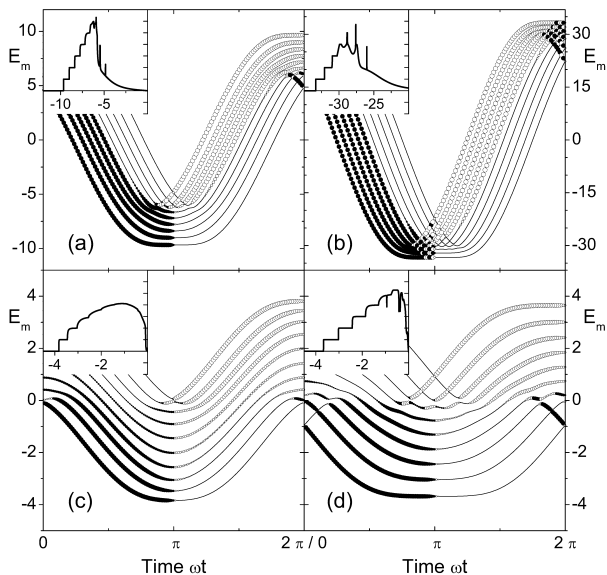}\\
  \caption{Instantaneous eigenenergies $E_m$ (in units of $J_d$)
  for $V_g=0$, $N=10$ and (a) $P=8 J_d$, $\lambda=4 L$; (b) $P=32
  J_d$, $\lambda=4 L$; (c) $P=2 J_d$, $\lambda=4 L$; and  (d) $P=2
  J_d$, $\lambda=2 L$. Insets: the pumped charge $Q^{\text{res}}$
  as  function of the scaled gate voltage $V_g+E_F$; the distance
  between the ticks on the ordinate axis is equal to a unit
  charge. }\label{fig3}
\end{figure}

Despite a certain inherent uncertainty of our estimates, they
prove useful for understanding the effect of changing the
amplitude and the wavelength of the SAW (Fig.~\ref{fig3}).
In Fig.~\ref{fig3}a, the number of steps is close to optimal,
$N/2=5$, and is limited by the localization criterion
$N_\text{steps} \approx N_1=3.9$. Increasing $P$ by a factor of 4
(Fig.~\ref{fig3}b) reduces the number of steps due to discrete
lattice effects: $N_\text{steps}\approx N_2=3.2$. One can clearly
see that for higher energy levels (close to the band center) the
tight-binding coupling $J_d$ is no longer relevant: $E_m(t)$ with
$m > N_2$ follow a sequence of cosine curves $P \cos(\omega t +
\delta \phi)$ with equal phase differences $qa$. These curves
correspond to the individual on-site energies $\epsilon_n(t)$. In
this regime the hopping amplitude $J_d$ leads only to tiny
anti-crossings between the energy levels, which in turn give rise
to the sharp peak-like structure in the pumping curve. The effects
of the tight-binding approximation become less pronounced as $P$
is reduced below the optimum (Fig.~\ref{fig3}c). In this case
$N_\text{steps}\approx N_1=1.96$ and the peaks in the pumping
curve are suppressed. The missing steps can be brought back by
shortening the wave-length, as shown in Fig.~\ref{fig3}d. The
estimated number of steps is now the same as in the original case
(a). However, the non-parabolic shape of the potential minimum is
more pronounced. Note that in case (d) the flat region for
$E_0(t)$ extends over half of the period, since $\lambda = 2L$.

\section{Perturbations of the pumping potential}\label{SecPerturbations}
The pumping potential in Eq.~\eqref{potential_original} is of a
rather high symmetry. Small perturbations -- such as a static
impurity or a reflected SAW --- can change the shape and the
position of the current quantization steps. In order to explore
these effects, we  add to $\epsilon_n(t)$  a smooth function of
$x_n$ and $t$,
\begin{align} \label{potGenreallyPerturbed}
  \epsilon_n(t) & = - V_g + P \cos(\omega t - q x_n)
  + U(x_n, t) \, .
\end{align}
Similarly to the situation discussed above, the structure of the
relevant energy levels can be analyzed using the harmonic
approximation, provided that $U(x_n, t)$ changes slowly and the
travelling wave-packet is well localized: $\xi_m \, \partial
U(x,t)/\partial x \ll \Delta$. The first-order approximation for
the instantaneous energy,
\begin{align}
  E_m(t)= -P-2 J_d + \Delta (m -1/2) + U(x_0(t),t) \, ,
\end{align}
is valid for $|x_0(t)| \lesssim \frac{L}{2} - \xi_m$ (we consider
the case $\lambda > L$). Note that $x_0(t)$ is the position of the
potential well minimum, and $x_0(t_0)=0$ (the middle of the
channel). Now even within the harmonic approximation $E_m(t)$ is
explicitly time-dependent and this time dependence maps onto the
shape of the current quantization steps. To make a quantitative
statement we note that the instantaneous wave-function remains
unperturbed in  first-order; $|\psi_N(t)|$ becomes greater than
$|\psi_1(t)|$ at $t=t_0$. At this point, the partial decay widths
are equal, $\Gamma_l^{(m)}=\Gamma_r^{(m)}$, and the resonance
approximation yields a half-integer pumped charge. Therefore, the
transition between the consecutive plateaus takes place at the
gate voltages $V_g^{(m)}=E_m(t_0) - E_F$. In particular, half of
the first step in the pumping curve is reached at the gate voltage
\begin{align} \label{V12general}
  V_{1/2} \equiv V_g^{(1)} = V_0 + U(0, t_0) \ ,
\end{align}
such that $Q(V_{1/2})=e /2$. Here  $V_0 =-E_F-P -2 J_d + \Delta/2$
is the threshold voltage for the first step in the absence of
perturbations.

The resonance moment associated with the left-right transition at
$V_g=V_{1/2}$ is well defined, since the energy levels  $E_m(t)$
are in general no longer constant in the vicinity of $t=t_0$.
Therefore, the slope of the first quantization step can be
estimated from the resonance approximation.  The value of the
total pumped charge at $V_g=V_{1/2}+\delta V$ is dominated by the
unloading resonance at $t=t_0+\delta t$, where $\delta V =
\dot{E}_1(t_0) \delta t$. The other resonances contribute charges
exponentially close to an integer; for simplicity, let us consider
only one loading through the left lead (which gives $\Delta Q_1
\approx e =\text{const.}$) before unloading at $t_0$. The
contribution of the latter, $\Delta Q_2(t)$, can be calculated
using the Gaussian wave-function \eqref{ground0} in
Eqs.~\eqref{decayWidth} and \eqref{Qrespartial}. The resulting
total pumped charge $\Delta Q_1 + \Delta Q_2$ is
\begin{gather}
 Q \approx e- \frac{ e \lvert \psi_1^{(1)}(t) \rvert^2 }{\lvert
 \psi_1^{(1)}(t)\rvert^2 +\lvert \psi_N^{(1)}(t) \rvert^2} =
 \frac{e}{2} \left ( 1+ \tanh \frac{L v \delta t}{\xi^2} \right )
 \, .
\end{gather}
We define the steepness of the first step, $S$, as
\begin{gather}
  S \equiv \frac{dQ}{dV_g}\Bigr |_{V_{g}=V_{1/2}}   \approx
  \frac{eLv}{\xi^2} \left [ \left| v \frac{\partial U}{\partial x}
  + \frac{\partial U}{\partial t}  \right |^{-1} \right
  ]_{\begin{subarray}{l}
  t  =t_0 \\
  x  =0 \\
  \end{subarray}} \, .
 \label{steepnessGeneral}
\end{gather}
The pre-factor in Eq.~\eqref{steepnessGeneral} is the least
accurate, since the applicability of Eq.~\eqref{ground0}  at the
ends of the channel is marginal. Taking the absolute value in
Eq.~\eqref{steepnessGeneral} makes the result valid for both signs
of $\dot{E}_1$ at $t=t_0$. Our derivation is not justified for
perturbations that yield small values of the denominator in
Eq.~\eqref{steepnessGeneral}. Then, the steepness remains bounded
due to the finite resonance width.

The quantization accuracy can be estimated along similar lines.
However, the results are less transparent since the energy levels
involved are beyond the simple harmonic approximation.

\subsection{Sensitivity to the second SAW} \label{SecSteepness}
For a particular example of a perturbation which mimics the
experimental situation, consider the following potential
\begin{align}
  U(x_n, t) & = P^{-} \, \cos \left( \omega  t + q x_n  + \varphi \right) +
  b \, x_n / L \, . \label{Uparticular}
\end{align}
Here $P^{-}$ and $\varphi$ are the amplitude and the phase of a
second SAW, propagating in the negative direction. It can be
generated either due to reflections of the main beam
\cite{Talyanskii97} or by a second transducer \cite{Cunningham99}.
We also include a simple static perturbation [proportional to $b$
in Eq.~\eqref{Uparticular}] which breaks the left-right symmetry
of the channel in the absence of the SAW. The estimates in
Eqs.~\eqref{V12general} and \eqref{steepnessGeneral} become
\begin{gather}
  V_{1/2}  = V_0 - P^{-} \cos{\varphi} \, , \label{V12particular} \\
  S  = \frac{ 4 N_1 e}{|b+ 2 q L P^{-} \sin \varphi|} \, . \label{Sparticular}
\end{gather}
[We have used the relation $N_1=L^2/(4\xi^2)$ in
Eq.~\eqref{Sparticular}.]

\begin{figure}[tb]
  \includegraphics[width=8cm]{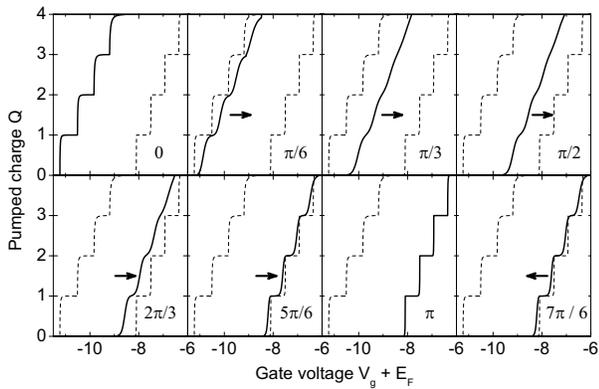}\\
  \caption{Interference of two counter-propagating SAWs with an amplitude ratio
   $P^{-}/P=0.2$. The phase difference $\varphi$ is changed from $0$
  to $7 \pi/6$, in steps of $\pi /6$, as indicated. Solid curves
  show the pumped charge (in units of $e$) versus the gate
  voltage (in units of $J_d$), the dotted lines mark the best
  quantization conditions achieved at $\varphi=0$ and
  $\varphi=\pi$. For $\varphi \in [\pi, 2 \pi]$ the pumping curves
  repeat the same sequence in reverse order (not shown). The
  parameters used are: $P=8 J_d$, $\lambda=4 L$, $N=10$. Curves are computed
  using the resonance approximation. }\label{phi_a}
\end{figure}
First we consider the case of a reflected wave only ($b=0$). A
series of pumping curves for different values of the phase
difference is presented in Fig.~\ref{phi_a}. As can be seen from
Eq.~\eqref{V12particular}, the threshold voltage changes
periodically in $\varphi$, reaching extremal values at $\varphi=0$
and $\pi$. Between these special values of $\varphi$, the
staircase structure is more smooth,  and the steps are more
symmetric: the convex and the concave parts of a step become
almost congruent. The pumping curves are identical for $\pm
\varphi$ due to the symmetry of the potential
\eqref{potGenreallyPerturbed} with $b=0$.

\begin{figure}[tb]
  \includegraphics[width=8cm]{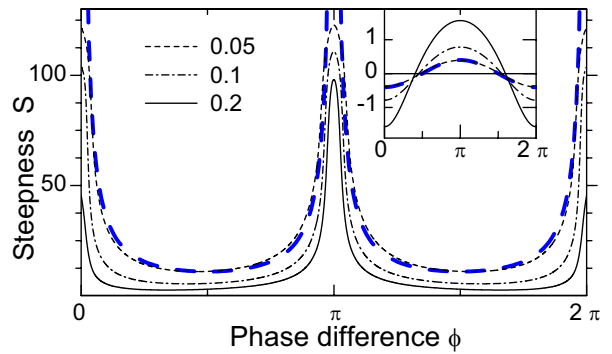}\\
  \protect\caption{(Color online) The steepness of the first step
  $S=d Q/d V_{g}$ at $Q=e(1/2)$, in units of $e \, J_d^{-1}$ for
  $P^{-}/P= 0.05, 0.1, 0.2$ as  function of the phase difference
  $\varphi$.  The curve for $P^{-}/P=0.2$ corresponds to the data
  presented in Fig.~\ref{phi_a}. Inset: threshold voltage
  $V_{1/2}(\varphi)-V_0$ versus $\varphi$, in units of $J_d$.
  Curves are computed exactly from Eq.~\eqref{chargeexact}. Thick
  dashed (blue) lines show analytic estimates, given by
  Eqs.~(\ref{V12particular}--\ref{Sparticular}) for the smallest
  amplitude ratio $P^{-}/P= 0.05$; the pre-factor $N_1$ in
  Eq.~\eqref{Sparticular} has been treated as a free fitting
  parameter. The parameters used are: $P=8 J_d$, $\lambda=4 L$,
  $N=10$,  $ka=\pi/5$, $\tilde{J}=1$. }\label{refsym}
\end{figure}
For a quantitative characterization of the second SAW effect we
have determined numerically the positions and the slope of the
pumping curves at $Q=e/2$ without any approximations in
Eq.~\eqref{chargeexact}. The results are shown in
Fig.~\ref{refsym}. Tuning the phase difference $\varphi$ for a
fixed amplitude ratio $P^{-}/P$ to the values at which the r.h.s.
of Eq.~\eqref{Sparticular} diverges enhances the steepness of the
first step by orders of magnitude. The sharpest steps are achieved
at the extrema of the threshold voltage $V_{1/2}$, as shown in the
inset in Fig.~\ref{refsym} and qualitatively in Fig.~\ref{phi_a}.

The above example shows that a symmetric pumping potential is
favorable for quantization: the steepest plateaus are achieved
without a secondary SAW or with $P^{-} \not =0$ and $\varphi=0$,
$\pi$, when the total SAW potential $V(x_n,t)\equiv\epsilon_n(t)$
is invariant  under $x_n \rightarrow -x_n$, $t \rightarrow
-t+\text{const}$.

Further reduction of symmetry is achieved by choosing $b\not = 0$
and $P^{-}\not = 0$. Here two regimes are possible. For small $b$,
the situation is similar to the previous case: the steepness is
greatly enhanced at two values of $\varphi$ between $0$ and
$2\pi$,  when the denominator in Eq.~\eqref{Sparticular} vanishes.
In contrast, for $b> 2 q L P^{-}$ it is the static asymmetry of
the channel that determines the slope of the steps, which now has
only one wide maximum as function of $\varphi$.
\begin{figure}
  \includegraphics[width=8cm]{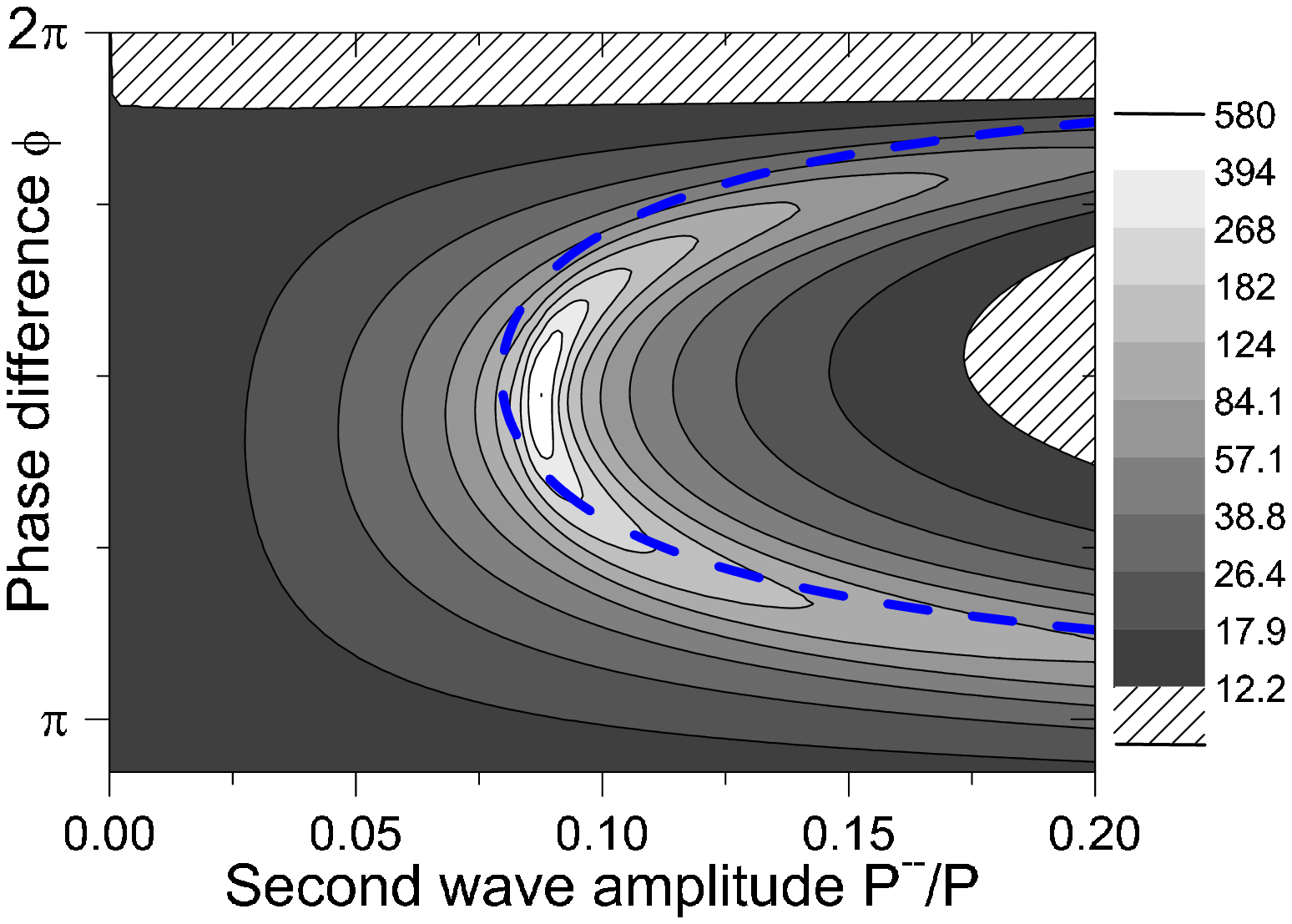}\\
  \caption{(Color online) The steepness of the first step for
   $P= 2 J_d$, $\lambda= 2 L$, $N=10$, $b=1 J_d$, as  function of
  $P^{-}/P$ and $\varphi$. The dashed (blue) line marks the
  combinations of amplitude and phase at which the first order
  estimate [Eq.~\eqref{Sparticular}] diverges. Note the
  logarithmic grey-coding scale. The steepness without the
  perturbation ($P^{-}=b=0$) is $S=613 \, e/J_d$.
  }\label{refgrayscale}
\end{figure}
This behavior is illustrated in Fig.~\ref{refgrayscale}, which
shows the slope of the first step  as a function of the second SAW
amplitude and phase. The initial steepness at $P^{-}=0$, $b=1J_d$
is $S=12.2 (e/J_d)$ for the selected model parameters. That is
more than an order of magnitude less than in the unperturbed
($P^{-}=b=0$) case. Increasing the amplitude of the second SAW
improves the steepness for $\pi \lesssim \varphi \lesssim 2 \pi$
with a single wide maximum at $\varphi \approx 3 \pi/2$, in
agreement with Eq.~\eqref{Sparticular}. At $P^{-} \approx b/(2 q
L)$ the steepness peaks sharply, almost reaching the unperturbed
value. Further increase of $P^{-}$ reduces the steepness
gradually, which now has two maxima in $\varphi$, approaching
$\varphi=\pi$ and $\varphi=2  \pi$ for large $P^{-}$, similarly to
Fig.~\ref{refsym}. This example shows that a weak
counter-propagating  SAW with properly chosen amplitude and phase
is able to compensate for the static asymmetry of the pumping
potential and significantly improves the sharpness of the current
quantization steps.

Available experimental data are consistent with our conclusions.
Periodic oscillations of $V_{1/2}$ have been observed in
experiments with one active SAW transducer when the frequency of
the SAW was varied. The period of these oscillations was found to
match a full $2 \pi$ phase shift between the main SAW and a weak
wave reflected from the other (inactive) transducer
\cite{Talyanskii97}. Later experiments, with two active
transducers on both sides of the constriction, have confirmed this
scenario, and a sequence of pumping curves similar to our
Fig.~\ref{phi_a} has been reported \cite{Cunningham99}. Tuning of
the second SAW amplitude and phase has enabled the authors of
Ref.\ ~\onlinecite{Cunningham99} to improve the flatness of the
first quantization plateau.

The key argument leading to Eq.~\eqref{Sparticular} concerns the
gate voltage dependence of the potential profile at the point
where capture/release of an electron happens with equal
probabilities from either side of the barrier. Therefore, the
phase and amplitude dependence of the steepness,
$S(\varphi,P^{-})$, is expected to be insensitive to the
particular choice of the pumping potential, as long as it leads to
a clear sequence of current quantization steps. We suggest the
following generic scenario of the plateau quality improvement,
that can be checked by detailed measurements using existing
experimental setups. One should measure the traces of the first
step steepness $S(\varphi)$ as function of the reflected wave
phase $\varphi$ for a set of gradually increasing secondary beam
amplitudes $P^{-}$. At small amplitudes,  $P^{-} < P^{-}_c$, the
steepness is expected to have one broad maximum at some
$\varphi=\varphi_0$. As $P^{-}$ is increased, the value at the
maximum, $S(\varphi_0)$, increases and at $P^{-}=P^{-}_c$, the
maximum splits into two, $S(\varphi_1)$ and $S(\varphi_2)$, with
$\varphi_{1,2} = \varphi_0 \pm \arccos(P^{-}_c/P^{-})$, as shown
in Fig.~\ref{refgrayscale} by the dashed line [for our model
calculation $\varphi_0=3 \pi /2$ and $P^{-}_c \approx b/(2 q L)$].

\subsection{Source-drain bias and variations of screening} \label{biasSec}
Experimentally, acoustoelectric current can be studied along the
full crossover, from the depleted to the transmissive state of the
quantum wire, by changing the voltage on the depleting gate. Our
discussion so far has been concentrated on the quantized
single-electron transport, which is observed in the depleted
regime. As the first conduction channel opens, the shape of the
pumping potential in real space as well as screening effects
become increasingly important \cite{Levinson00PRL} and the
usefulness of our simplified 1D spinless electron model is very
limited. Keeping these limitations in mind, we will choose model
parameters that most closely correspond to a point contact near
the depletion threshold, and illustrate the breakdown of quantized
transport.

For $P > J_d=\tilde{J}$, the tight-binding band is significantly
deformed  (see Fig.~\ref{fig3}), therefore we choose a relatively
small SAW amplitude $P= 0.5 J_d$, but a large number of sites
$N=24$ to maintain $N_{\text{steps}} >1$. The Fermi wave number
$ka=\pi/12$ is taken close to the band bottom.

\begin{figure*}[tb]
  \includegraphics[width=12cm]{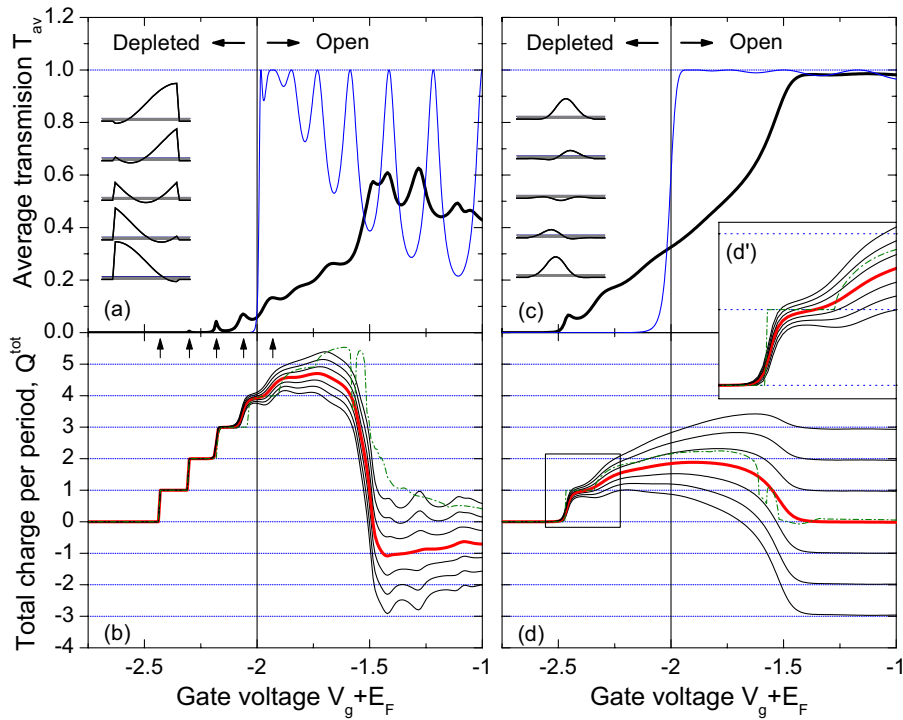}\\
  \caption{(Color online) Crossover between the depleted and open regimes of the
  conduction channel for $N=24$, $ka = \pi/12$, $\tilde{J}=J_d$. (a) Time-averaged
  transmission coefficient $\mathcal{T}_\text{av}$ in the absence [thin
  (blue) line] and presence (thick black line) of a SAW with
  $P=0.5 J_d$, $ \lambda= 2 L$. (b) The pumped charge $Q$ [thick
  (red) line] and the total charge $Q^{\text{tot}}$ (thin black lines)
  for bias voltages
  $e V_{SD}/(\hbar \omega)$ changing from $-3$ to $3$ in steps of
  one. The dashed-dotted line shows $Q^{\text{res}}$ at no bias [resonance approximation,
  Eq.~\eqref{Qtot}]. (c,d) The same as in
  (a,b), but with an exponentially screened pumping potential
  [Eq.~\eqref{screenedpot}]. A sequence of snap-shots in panels
  (a,b) and (c,d) shows the corresponding pumping potential in
  real space for $V_g$ in the middle of the first plateau; time
  increases from top  to bottom, $\omega t \in [\omega t_0- \pi/2
  \omega; \omega t_0+ \pi/2 ]$. }\label{bias}
\end{figure*}

Consider first the situation before the SAW is applied ($P=0$).
The zero-bias dc conductance of the channel is determined by the
transmission coefficient $\mathcal{T}$ (Landauer formula, see
Sec.~\ref{FullCalc}). There is a potential barrier between the
left and the right reservoirs for $-V_g>0$, therefore the value of
$V_g + E_F =-2 J_d \cos ka \approx -2 J_d$ is expected to be the
borderline between transmissive and blocked states of our channel.
This corresponds to the depletion threshold of a true point
contact. We plot the transmission coefficient in the
\emph{absence} of SAW versus gate voltage, $\mathcal{T}(V_g)$,  in
Fig.~\ref{bias}a with a thin (blue) line. For $V_g<0$, the
transmission is exponentially blocked by a rectangular barrier of
height $\approx -V_g$ and length $L$, while above the depletion
threshold a Fabry-Perot-like pattern of high transmission is
observed due to multiple reflections at the sharp ends of the
constriction.

At a non-zero SAW amplitude, the transmission coefficient
$\mathcal{T}(t)$ becomes time-dependent and the adiabatic formula
[Eq.~\eqref{bias_current}] should be used to relate it to the
conductance. In the linear response regime, the second term in the
curly brackets in Eq.~\eqref{bias_current} is proportional to
$\partial^2 f/\partial E^2$ and can be neglected
\cite{Entin02form}. This results in a generalized Landauer formula
\cite{Entin02form}, $G=(e^2/h) \, \mathcal{T}_{\text{av}}$, where
$\mathcal{T}_{\text{av}} \equiv (\omega/2 \pi) \int_0^{2
\pi/\omega} \!\! dt \, \mathcal{T}(t)$ is the time average of the
instantaneous transmission coefficient $\mathcal{T}(t)$. This
quantity is plotted in Fig.~\ref{bias}a with a thick black line.
One can see that switching on the SAW smears the sharp step in the
conductance over the range of $\pm P$ around the depletion
threshold. Qualitatively similar smoothing of the conductance
quantization steps due to SAW  has been observed experimentally
\cite{Shilton96,Talyanskii97}.

Figure~\ref{bias}a also shows some additional structure below the
depletion threshold. This structure is correlated with the pumping
curve shown by a thick (red) line in Fig.~\ref{bias}b. Comparing
$Q(V_g)$ and $\mathcal{T}_\text{av}(V_g)$ we see that each step in
the acoustoelectric current is associated with a peak in the time
averaged transmission as indicated by arrows in Fig.~\ref{bias}a
(the first two peaks are too small to be seen on a linear scale).
It is easy to explain the origin of these peaks using the
resonance approximation diagram (see Fig.~\ref{fig2}). At gate
voltages between the quantization plateaus the system remains at
resonant transmission for a considerable fraction of the period,
therefore $\mathcal{T}_\text{av}$ becomes greatly enhanced.

In the presence of both SAW and source-drain bias, the total
charge transfer per period, $Q^{\text{tot}} \equiv \int_0^{2
\pi/\omega} \!\! dt \, I_l(t)$, becomes
\begin{equation} \label{EqQtot}
  Q^{\text{tot}} = Q + (e^2 V_{SD}/\hbar \omega) \, \mathcal{T}_{\text{av}}
\end{equation}
(in the linear response regime). The result is a sum of the two
terms: pure pumping contribution [thick (red) curve in
Fig.~\ref{bias}b]; and the average transmission (thick curve in
Fig.~\ref{bias}a), multiplied by a constant. Equation
\eqref{EqQtot} suggests that $\hbar \omega$ is a natural unit for
the source-drain energy mismatch $e V_{SD}$. In the quantized
pumping region, the contribution of the bias, $I^{\text{bias}}$,
becomes comparable to that of pumping, $I^{\text{pump}}$, if the
bias voltage source transports several electrons per cycle. When
$\mathcal{T}_{\text{av}}$ is of order one, this regime is attained
for $e V $ equal to several $\hbar \omega$. Thus for a qualitative
picture of the pumping curve in the presence of bias, we have
plotted $Q^{\text{tot}}$ for the bias voltage $e V_{SD}/\hbar
\omega$ ranging from $-3$ to $3$ by thin black lines in
Fig.~\ref{bias}b. The main observation is that the higher is the
step number the more sensitive it is to the bias (as one can
already appreciate form the average transmission curve). Similar
behavior is reported in experimental studies
\cite{Shilton96,Talyanskii97}.

The main results of the above discussion remain unchanged if a
phenomenological screening \cite{Levinson00PRL,Maksym00} is
introduced:
\begin{align}\label{screenedpot}
  \epsilon_n(t) = \left [ -V_g+
  P \, \cos \left( \omega  t - q x_n  \right) \right ] \exp
  \left(-x_n^2/L_s^2 \right ) \, .
\end{align}
We have repeated the previous calculation using the same values of
parameters but modified the pumping potential \eqref{screenedpot}
with $L_s=L/4=\lambda/8$. The results are shown in
Figs.~\ref{bias}c and \ref{bias}d. The main qualitative difference
is the disappearance of the interference pattern in the
transmission curve both with and without the SAW. The number of
steps is reduced to one (shown separately in
Fig.~\ref{bias}d${}^\prime$), since the effective amplitude of the
SAW is decreased by the screening factor in
Eq.~\eqref{screenedpot}. Calculations with larger values of $P$
produce more steps, along the same lines as discussed in
Sec.~\ref{subSecNumSteps} for the unscreened potential
\eqref{potential_original}. We have also checked that the behavior
of the first step steepness $S(P^{-},\varphi)$ as function of
reflected  SAW amplitude and phase, follows the general scenario
suggested in Sec.~\ref{SecSteepness}.

We note that in the above example (Fig.~\ref{bias}) the resonance
approximation still holds below the depletion threshold, when a
moving quantum well is isolated from the Fermi sea in the leads.

\section{Discussion and conclusions} \label{SecDiscussion}
Quantized  electronic transport, driven by SAW's, has been
considered in several recent theoretical studies
\cite{Aizin98,Gumbs99,Flensberg99,Maksym00,Galperin01,
Robinson01,Margulis02}. Here we discuss our approach in relation
to those works.

Several of models \cite{Aizin98,Gumbs99,Flensberg99,Robinson01}
make a distinction between electrons \emph{already localized} in a
moving potential well (dynamic quantum dot) and  those belonging
to the Fermi sea. The current is then calculated by considering
the loss of electrons from the dynamic quantum dot at the stage of
its formation \cite{Flensberg99,Robinson01} and/or its subsequent
motion \cite{Aizin98,Gumbs99,Flensberg99,Robinson01}. This
approach presupposes the formation of the dynamic quantum dot, but
does not require it to be at thermodynamic equilibrium with the
reservoirs at all times. Moreover, all the quantization error
mechanisms within these models (gradual back-tunneling
\cite{Aizin98,Gumbs99,Flensberg99}, non-adiabaticity at the
formation stage \cite{Flensberg99} and non-equilibrium dynamics
during the transfer \cite{Robinson01}{}) consider electrons with
energies that can significantly exceed the Fermi energy in the
remote reservoirs.

Our adiabatic quantum calculation \cite{Aharony02PRL} differs from
these studies  in two significant aspects: (i) The formation of a
dynamic quantum dot is not a necessary condition for the
calculation of the acoustoelectric current. We do identify,
however, the localized electronic states (whenever such states are
present) via the resonance approximation and confirm that they are
responsible for the quantized transport. (ii) In the adiabatic
pumping approximation \cite{Entin02form}, the time-dependent
potential never excites the carrier by a finite amount of energy
away from the Fermi level \cite{MoskaletsButtiker02}. Therefore,
we never observe quantization steps when the moving potential well
rises above the Fermi level upon passing through the middle of the
channel.

We find the numerical calculations by Maksym \cite{Maksym00} to be
the closest to our study. He considers a 1D single-particle model
with a pumping potential similar to our Eq.~\eqref{screenedpot}.
The current at the quantization plateaus is found to be carried by
the lowest energy states of the local potential minimum, in
accordance with our results.

The  quantization accuracy in our approach is determined by two
factors which we expect to become experimentally relevant for
sufficiently low tunnelling barriers. The first one is the
possibility of \emph{both} reservoirs to participate in the
capture/release of an electron. This error mechanism is covered by
the resonance approximation and leads to simple estimates like our
formulas for the first step steepness discussed in
Sec.~\ref{SecSteepness}. The second factor concerns mixing of the
localized states with continuum in the leads, which can give
significant width to the quasi-stationary states in the moving
quantum dot. Compared to the predictions of the loading/unloading
scenario (Sec.~\ref{ResonanceCalc}), this effect further degrades
the flatness of the quantization  (see, e.g., Figs.~\ref{fig1}b,
\ref{refsym} and \ref{bias}d) and eventually leads to the
breakdown of the quantized transport as the channel opens
(Sec.~\ref{biasSec}).

We have not considered explicit Coulomb interactions between
electrons in the depleted part of the channel, which set the
energy scale of the problem. One can make a naive estimate of the
level spacing $\Delta$, which in the continuous limit is
 \cite{Aharony02PRL} $\Delta =\hbar q \sqrt{P/m^{\ast}}$. Using
typical experimental values \cite{Cunningham99} for the SAW
amplitude $P=20 \text{meV}$, wavelength $\lambda = 1 \mu\text{m}$
and GaAs bulk effective mass $m^{\star}=0.067 m_0$, one gets
$\Delta = 1 \text{meV}$, which is an order of magnitude less that
the distance between the quantization steps observed in
experiments  \cite{Shilton96, Talyanskii97,Cunningham99}. This
discrepancy can be qualitatively understood on a mean-field level:
if an electron is captured by the moving potential minimum, its
unscreened electric field makes the potential well seen by the
other electrons much shallower, and thus increases the spacing
$\Delta$ between resonances by the amount of the charging
energy \cite{Moskalets01,Berkovits03}. Such a picture is also
supported by the numerical calculation of a two-electron problem
by Gumbs and co-workers \cite{Gumbs99}. It is plausible that the
effective values for the parameters of our model can be estimated
from a self-consistent realistic calculation.

In conclusion, we have considered a simple model for SAW-driven
adiabatic pumping of electrons through a quasi-1D quantum wire. A
stair-case structure of the acoustoelectric current has been
mapped onto the instantaneous energy spectrum of the pumping
potential. Numerical calculations and analytic estimates confirm
the experimentally observed behavior of the acoustoelectric
current as function of the SAW amplitude, wavelength, source-drain
bias, and the parameters of a weak counter-propagating beam.
Quantitative measurements of the plateau quality as a function of
the second SAW amplitude and phase are proposed to probe the
relevance of our model. The presented single-electron picture
captures all the main features of the quantized transport.
\begin{acknowledgments}
  This project was carried out in a center of excellence supported
  by the Israel Science Foundation.
\end{acknowledgments}


\end{document}